\documentclass[aps,twocolumn,pra,
preprintnumbers,amsmath,amssymb,superscriptaddress]{revtex4}
\usepackage{graphicx}
\usepackage{color}

\definecolor{refkey}{rgb}{0.9451,0.2706,0.4941}
\definecolor{labelkey}{rgb}{0.9451,0.2706,0.4941}
 
\newcommand{\eq} {equation}
\newcommand{\eqa} {eqnarray}
\newcommand{\NN} {\nonumber}
\newcommand{\eps} {\epsilon}
\newcommand{\bpsi} {\bar{\psi}}

\newcommand{\vtheta}{\vartheta}
\newcommand{\sgn}{{\rm sgn}}

\newcommand{\beps} {\bar{\epsilon}}
\newcommand{\blam} {\bar{\lambda}}

\newcommand{\gam} {\gamma}

\def\del{\partial}
\begin{document}

\preprint{WIS/03/17-Oct-DPPA}

\title{
Supersymmetric solutions and Borel singularities 
for $N=2$ supersymmetric Chern-Simons theories
}

\date{\today}

\author{Masazumi Honda}\email[]{masazumi.hondaATweizmann.ac.il} 
\affiliation{{\it Department of Particle Physics, Weizmann Institute of Science, Rehovot 7610001, Israel}}

\begin{abstract}
In supersymmetric (SUSY) field theory,
there exist configurations
which formally satisfy SUSY conditions
but are not on original path integral contour.
We refer to such configurations as complexified supersymmetric solutions (CSS).
In this paper
we discuss that
CSS provide
important information on large order behavior 
of weak coupling perturbative series in SUSY field theories.
We conjecture that
CSS with a bosonic (fermionic) free parameter
give poles (zeroes) of Borel transformation of perturbative series
whose locations are uniquely determined by actions of the solutions.
We demonstrate this 
for various SUSY observables 
in 3d $\mathcal{N}=2$ SUSY Chern-Simons matter theories on sphere.
First we construct infinite number of 
CSS in general 3d $\mathcal{N}=2$ SUSY theory with Lagrangian
where adjoint scalar in vector multiplet takes a complex value 
and matter fields are nontrivial. 
Then we compare their actions 
with Borel transformations of perturbative expansions by inverse Chern-Simons levels
for the observables
and see agreement with our conjecture.
It turns out that
the CSS explain all the Borel singularities for this case.
\end{abstract}

\maketitle


\noindent

\section{Introduction}
Supersymmetric (SUSY) field theories
have provided good laboratories to find and sharpen ideas in theoretical physics.
In SUSY field theories,
configurations preserving a part of SUSY play prominent roles
because of their stabilities,
restricted quantum corrections and so on.
While most of attention has been paid
to SUSY solutions which take values on original path integral contour,
there also exist configurations
which formally satisfy SUSY conditions
but are not on the original contour.
We refer to such configurations as complexified supersymmetric solutions (CSS).
Compared to standard SUSY solutions,
CSS have been much less appreciated. 
Purpose of this paper is to point out a physical role of CSS.

Since SUSY conditions are often sufficient to satisfy saddle point equation,
CSS are also often complex saddle points
which are saddle points but not on original path integral contour.
Therefore it is worth to note historical situations 
on complex saddle points in quantum field theory (QFT).
While saddle points along original contour always contribute to path integral,
complex saddles may or may not contribute
since this is determined by details 
on steepest descents associated with them.
Historically most physicists have not taken this possibility into account
in saddle point analysis of path integral
despite its importance was pointed out long time ago \cite{Balian:1978et}.
Namely,
people have usually considered only saddles on the original contour 
and not discussed whether or not complex saddles contribute.
Recently there appeared relatively more examples in the context of resurgence 
where some complex saddles actually give important contributions
to path integral,
and
are necessary to obtain unambiguous answer from resummation of weak coupling perturbative series \cite{Cherman:2014ofa}.
Although the examples are much simpler than typical interacting QFT, 
which are either quantum mechanics, two dimensional or topological,
there is no a priori reason to doubt that 
similar things happen in more general setup.
However, 
it is usually difficult to check this explicitly
since we need to construct complex saddles
and find large order behaviors of perturbative series around all the contributing saddles.
Therefore SUSY field theories would provide
good next steps to understand roles of complex saddles in QFT.

The above facts suggest that
CSS may importantly contribute to path integral.
Indeed recent works on SUSY localization \cite{Pestun:2007rz} imply that
CSS sometimes give important contributions
\cite{Fujitsuka:2013fga,Benini:2013yva,Benini:2015noa}.  
In this paper 
we do not discuss 
whether CSS are contributing saddles or not.
Here we propose that
independently of this question,
CSS provide
important information on large order behavior 
of weak coupling perturbative series in SUSY field theories.

Perturbative series in QFT is typically non-convergent \cite{Dyson:1952tj}.
One of standard ways to resum non-convergent series
is Borel resummation.
Given a formal series 
$P(g)=\sum_{\ell =0}^\infty c_\ell g^{a+\ell} $,
its Borel resummation along 
the direction $\theta$ is defined by
\begin{\eq}
\mathcal{S}_\theta P (g)
=\int_0^{\infty e^{i\theta}} dt\ 
e^{-\frac{t}{g}} \mathcal{B}P(t) ,
\label{eq:Borel} 
\end{\eq}
where $\mathcal{B}P(t)$ is analytic continuation of 
the formal Borel transformation  
$\sum_{\ell =0}^\infty 
\frac{c_\ell}{\Gamma (a+\ell )} t^{a+\ell -1}$.
In general Borel transformation have singularities 
in complex $t$-plane
and their locations contain important information.
Technically they determine
whether or not the perturbative series is Borel summable given $\theta$,
and
large order behavior of perturbative series
as they give radius of convergence of the Borel transformation.
Physically, according to Lipatov's argument \cite{Lipatov:1976ny},
Borel singularities are interpreted as non-trivial saddle points
and their actions determine locations of the singularities. 

In this paper
we conjecture 
a relation between CSS and
analytic properties of Borel transformation in SUSY field theories.
Let us consider
weak coupling perturbative expansion $P(g)$ around a saddle point
with certain topological numbers.
In general
there exist two types of CSS
with the same topological numbers:
one has a bosonic parameter while the other has a fermionic one.
We refer to
them as bosonic and fermionic complexified supersymmetric solutions, respectively.
We propose that
bosonic CSS
basically give poles of Borel transformation of $P(g)$
while fermionic CSS give zeroes.
More precisely
if there are $n_B$ bosonic and $n_F$ fermionic solutions
with the topological number 
and the action $S=S_c /g$,
then we conjecture that 
the Borel transformation 
has a pole for $n_B \geq n_F$ and zero for $n_B \leq n_F$ with degree $|n_B -n_F |$ at $t=S_c$.
Namely 
the Borel transformation includes the following factor 
\begin{\eq}
\mathcal{B}P^{(I)}(t)
\supset \prod_{\rm solutions} \frac{1}{(t-S_c )^{n_B -n_F}} .
\end{\eq}
Note that
CSS gives information not only on locations of singularities
but also on their degrees.
This leads us to further insights 
into non-perturbative structures of path integrals.

In next sections
we demonstrate the above story for various SUSY observables 
in 3d $\mathcal{N}=2$ SUSY Chern-Simons (CS) matter theory 
on sphere \footnote{
We consider theories with well-defined sphere partition functions
though ill-defined cases are also interesting \cite{Morita:2011cs}.
}.
In sec.~\ref{sec:construction}
we construct
infinite number of CSS 
in general 3d $\mathcal{N}=2$ theory with Lagrangian.
They are described by zero-modes of differential operators appearing in SUSY transformations of chiral multiplets
\footnote{
It was pointed out in \cite{Aniceto:2014hoa} that
for some rank-1 theories,
Borel singularities come from zero-modes of kinetic terms.
Our CSS are also a part of them.
}.
We cannot have such zero-modes for most case
when adjoint scalar $\sigma$ in vector multiplet is real 
i.e. on the original path integral contour,
but relaxing the reality condition allows such zero-modes.
In sec.~\ref{sec:analytic}
we compare 
their actions
with explicit Borel transformation 
found in \cite{Honda:2016vmv} 
(see also \cite{Russo:2012kj,Aniceto:2014hoa})
and see agreement with our conjecture.

\section{Construction of complexified supersymmetric solutions}
\label{sec:construction}
Let us construct
complexified supersymmetric solutions
in general 3d $\mathcal{N}=2$ SUSY gauge theories 
with Langragians on a squashed $S^3$.
We choose the squashed $S^3$ to be ellipsoid $S_b^3$
defined as the hypersurface in $\mathbb{R}^4$:
$ (x_1^2 +x_2^2 ) +b^2 (x_3^2 +x_4^2 ) $
$=$
${\rm Const.}$.
The round $S^3$ corresponds to the limit $b\rightarrow 1$.
As we will see,
the relation between CSS and Borel singularities
are more transparent for generic $b$
rather than the $b\rightarrow 1$ case
because actions of many solutions become coincident 
as $b\rightarrow 1$.

\subsection{Vector multiplet}
First we find CSS 
for the 3d $\mathcal{N}=2$ vector multiplet,
which consists of
gauge field $A_\mu$, adjoint scalar $\sigma$, gaugino $(\lambda ,\blam )$
and auxiliary field $D$.
For this purpose,
we need only the SUSY transformations
\footnote{We follow notation of \cite{Fujitsuka:2013fga}.
}:
\begin{\eqa}
\delta A_\mu   &=& -\frac{1}{2} \blam \gam_\mu \eps -\frac{1}{2} \beps \gamma_\mu \lambda, \quad
\delta \sigma  = \frac{i}{2} \blam\eps +\frac{i}{2} \beps \lambda, \NN\\
\delta \lambda &=& \left( \frac{1}{2} \eps_{\mu\nu\rho}F^{\nu\rho} - D_\mu \sigma \right) \gam^\mu \eps -i D\eps 
 -\frac{i}{f(\vtheta)} \sigma \eps ,\NN\\
\delta \blam   &=& \left( \frac{1}{2} \eps_{\mu\nu\rho}F^{\nu\rho} + D_\mu \sigma  \right) \gam^\mu \beps +iD\beps 
+\frac{i}{f(\vtheta)} \sigma \beps ,\NN\\
\delta D       &=& \frac{1}{2}\beps \gam^\mu D_\mu \lambda 
                   -\frac{1}{2}D_\mu \blam \gam^\mu \eps 
                   -\frac{1}{2}[\beps \lambda ,\sigma ]
                   +\frac{1}{2}[\blam\eps ,\sigma ] \NN\\
   &&               -\frac{i}{4f(\vtheta)}(\beps \lambda +\blam \eps ), 
\label{eq:QV}
\end{\eqa}
where
we take the coordinate
$(x_1 ,x_2 ,x_3, x_4)$ $=$ 
$(\cos{\vtheta}\cos{\varphi},$
$\cos{\vartheta}\sin{\varphi},$
$\sin{\vartheta}\cos{\chi},$
$\sin{\vartheta}\sin{\chi})$
and
$f(\vartheta ) $
$=$
$\sqrt{b^{-2}\sin^2{\vtheta} +b^{2}\cos^2{\vtheta}}$.
Our SUSY condition
is simply vanishing of all the transformations \eqref{eq:QV}.
As a sufficient condition,
we can solve the SUSY condition by \footnote{
This is not most general solution.
We could relate $F_{\mu\nu}$ to $D$. 
This class of solution includes vortices \cite{Fujitsuka:2013fga,Benini:2013yva}. 
}
\begin{\eq}
F_{\mu\nu} =0 ,\quad \sigma ={\rm const.},\quad 
D= -\frac{\sigma}{f(\vtheta )}  ,\quad
\lambda =\bar{\lambda}=0 .
\label{eq:SUSYV}
\end{\eq}
This is just Coulomb branch solution 
for real $\sigma$ 
but the SUSY condition can be formally satisfied 
even for complex $\sigma$.
Next we find CSS for chiral multiplet
under the configuration \eqref{eq:SUSYV}.
It turns out that
$\sigma$ is fixed to a particular complex value
by SUSY condition for chiral multiplet.

\subsection{Chiral multiplet}
The 3d $\mathcal{N}=2$ chiral multiplet consists of
scalars $(\phi ,\bar{\phi} )$, fermions $(\psi ,\bar{\psi} )$
and auxiliary fields $(F,\bar{F})$,
whose SUSY transformations are
\begin{\eqa}
\delta \phi      
&=& i\beps \psi ,\quad
\delta \bar{\phi}
= i\eps \bpsi ,\NN\\
\delta \psi      
&=& - \gam^\mu \eps D_\mu \phi -\eps \sigma \phi
                       -\frac{i\Delta}{f(\vtheta)} \eps \phi +i\beps F ,\NN\\
\delta \bpsi     
&=& -\gam^\mu \beps D_\mu \bar{\phi} -\bar{\phi} \sigma \beps
                        -\frac{i\Delta}{f(\vtheta)} \bar{\phi} \beps   +i\bar{F}\eps ,\NN \\
\delta F         
&=& \eps (-\gam^\mu D_\mu \psi +\sigma \psi +\lambda \phi ) 
                     +\frac{i(2\Delta -1)}{2f(\vtheta)}  \eps \psi ,\NN\\
\delta \bar{F}    
&=& \beps (-\gam^\mu D_\mu \bar{\psi} +\bar{\psi}\sigma  -\bar{\phi}\blam ) 
                     +\frac{i(2\Delta -1)}{2f(\vtheta)}  \beps \bar{\psi}  ,
\end{\eqa}
where we have assigned 
the $U(1)_R$-charges $(-\Delta ,\Delta , 1-\Delta ,\Delta -1 ,2-\Delta ,\Delta -2)$
to $(\phi ,\bar{\phi}, \psi ,\bar{\psi} ,F ,\bar{F})$ respectively. 
Under the vector multiplet configuration \eqref{eq:SUSYV},
we find two types of CSS
where one of them has a bosonic parameter
while the other has a fermionic parameter.
Below we take $\sigma$ to be Cartan part by using the gauge symmetry.

\subsubsection{Bosonic solutions}
First we look for solutions
with trivial $(\psi ,\bar{\psi},F,\bar{F})$
but nontrivial $(\phi ,\bar{\phi})$, namely
\begin{\eq}
\psi=\bar{\psi}= F=\bar{F}=0 ,
\end{\eq}
which leads us to $\delta \phi =\delta \bar{\phi}=\delta F =\delta\bar{F}=0$.
A nontrivial condition for $\phi$ comes from $\delta \psi =0$.
Decomposing the fields into components associated with weight vector $\rho$ 
of gauge group representation,
the condition for the component $\phi^\rho$ is
\begin{\eq}
\gam^\mu \eps D_\mu \phi^\rho +\eps (\rho\cdot \sigma ) \phi^\rho                  
+\frac{i\Delta}{f(\vtheta)} \eps \phi^\rho =0 .
\end{\eq}
We can easily find 
solutions for this condition
because eigenvalue problem of the differential operator above
is already solved in \cite{Hama:2011ea}
in the context of SUSY localization of $S_b^3$ partition function.
In order to compute its one-loop determinant for the chiral multiplet, 
\cite{Hama:2011ea} studied
the eigenvalue problem
\begin{\eq}
\gam^\mu \eps D_\mu \Phi  +\eps (\rho \cdot \sigma ) \Phi  +\frac{i\Delta}{f(\vtheta)}\eps \Phi 
= M \epsilon \Phi ,
\end{\eq}
where $M$ is the eigenvalue 
given by
\begin{\eqa}
&& M =M_{m,n}=\rho \cdot \sigma  +i \left( mb +nb^{-1} +\frac{Q}{2}\Delta \right) ,\NN\\
&& m,n\in \mathbb{Z}_{\geq 0} ,\quad Q=b+b^{-1} .
\label{eq:eigenB}
\end{\eqa}
The product of the eigenvalues roughly gives ``denominator" of the one-loop determinant.
Our SUSY condition is satisfied for $M=0$
but real $\sigma$ cannot satisfy this 
for generic $(b,\Delta )$.
If we relax this,
denoting eigenmode with the eigenvalue $M_{m,n}$ by $\Phi_{m,n}$,
we have the SUSY solutions
\begin{\eq}
\phi = \Phi_{m,n} ,\quad
\rho \cdot \sigma 
=\rho \cdot \sigma_{m,n}^{\rm bos}
=-i \left( mb +nb^{-1} +\frac{Q}{2}\Delta \right) .
\label{eq:bose}
\end{\eq}
Note that 
these conditions 
do not uniquely determine value of $\phi$
since 
we have freedom to change an overall constant.

\subsubsection{Fermionic solutions}
Next we find solutions
with trivial $(\phi ,\bar{\phi},F,\bar{F})$
but nontrivial $(\psi ,\bar{\psi})$, namely
\begin{\eq}
\phi=\bar{\phi}= F=\bar{F}=0 ,
\end{\eq}
which gives $\delta\psi =\delta\bar{\psi}=0$.
Then nontrivial SUSY conditions for $\psi$
come from vanishing of $\delta \phi $ and $\delta F$:
\begin{\eq}
\beps \psi^\rho =0 ,\quad
\eps (-\gam^\mu D_\mu  +\rho\cdot\sigma   )\psi^\rho   +\frac{i(2\Delta -1)}{2f(\vtheta)} \eps \psi^\rho =0 .
\end{\eq}
Again we can easily find solution of this condition
because \cite{Hama:2011ea} also solved the eigenvalue problem
\begin{\eqa}
&& \eps (-\gam^\mu D_\mu \Psi +\rho\cdot\sigma \Psi  )   +\frac{i(2\Delta -1)}{2f(\vtheta)}  \eps \Psi 
= M\epsilon \Psi , \NN\\
&& M =M_{m,n}
=\rho\cdot \sigma -i \left( mb +nb^{-1} -\frac{Q(\Delta -2)}{2} \right) , \NN\\
\label{eq:eigenF}
\end{\eqa}
to compute ``numerator" of the one-loop determinant.
The SUSY condition is satisfied for $M=0$
but again we need to take complex $\sigma$.
Thus,
denoting eigenmode with the eigenvalue $M_{m,n}$ by $\Psi_{m,n}$,
we have the SUSY solutions
\begin{\eqa}
&& \beps \psi^\rho =0,\quad \psi^\rho =\Psi_{m,n},\NN\\
&& \rho\cdot\sigma 
=\rho\cdot\sigma_{m,n}^{\rm fer}
=i \left( mb +nb^{-1} -\frac{Q(\Delta -2)}{2} \right) .
\label{eq:fermi}
\end{\eqa}
 
So far 
we have focused on the particular component $\rho$ of single chiral multiplet.
In general we have various chiral multiplets with various representations
and each of those has multiple components associated with weights. 
We can always construct 
a CSS where only one component of a chiral multiplet is \eqref{eq:bose} or \eqref{eq:fermi} and all the others are trivial.
This solution forces $\sigma$ to be in a $({\rm rank}(G)-1)$-dimensional hypersurface in $\sigma$-space.
Hence $\sigma$ is not completely determined for ${\rm rank}(G)>1$
and we can also take nontrivial configuration \eqref{eq:bose} or \eqref{eq:fermi} for other components.
In general there are solutions with multiple nontrivial components 
unless $\sigma$ is ``overdetermined",
namely 
SUSY conditions set 
more than ${\rm rank}(G)$ linearly independent hypersurfaces in the $\sigma$-space.

\section{Complexified supersymmetric solutions and analytic property of Borel transformation}
\label{sec:analytic}
Let us consider
3d $\mathcal{N}=2$ SUSY CS matter theory 
coupled to chiral multiplets of representations 
$\{ \mathbf{R_a}\}$ with $R$-charges $\{ \Delta_a \}$.
We take gauge group to be a product of semi-simple gauge groups:
$G=G_1 \times\cdots\times G_n$ with CS levels $k_1 ,\cdots ,k_n$ respectively.
We are interested in
perturbative expansion by $g_p \propto 1/|k_p |$
and a relation between its Borel transformation and the CSS.

\subsection{Squashed sphere}
In \cite{Honda:2016vmv},
the author found 
explicit finite dimensional integral representations 
of Borel transformations for $S_b^3$ partition functions.
As we briefly review below,
the Borel transformations are somehow hidden in localization formula
given by \cite{Hama:2011ea,Imamura:2011wg}
\begin{\eq}
Z_{S_b^3}(g)
=  \int_{-\infty}^\infty d^{{\rm rank}(G)} \sigma\
Z_{\rm cl}(\sigma ) Z_{\rm 1loop}(\sigma )  ,
\label{eq:ZS3}
\end{\eq}
where 
$Z_{\rm cl}$ is the classical contribution
and
$Z_{\rm 1loop}$ is the one-loop determinant in the localization procedure
\footnote{
$Z_{S_b^3}$ is independent of 
Yang-Mills couplings.
If we include Fayet-Illiopoulos term,
we have a linear function of $\sigma$ in 
$Z_{\rm cl}$ 
and this does not affect our results.
Real mass gives a constant shift of $\sigma$ in $Z_{\rm 1loop}$.
Correspondingly,
the CSS also have the same constant shift of $\sigma$.
}:
\begin{\eqa}
&& Z_{\rm cl}(\sigma )
= \exp{\Bigl[  \sum_{p=1}^n \frac{i\sgn (k_p )}{g_p} {\rm tr} (\sigma^{(p)} )^2  \Bigr]} ,\NN\\
&& Z_{\rm 1loop}(\sigma ) 
= \frac{\prod_{\alpha \in {\rm root}_+ }
 4\sinh{(\pi b \alpha \cdot \sigma )} \sinh{(\pi b^{-1} \alpha \cdot \sigma )}  }
   {\prod_a \prod_{\rho_a \in \mathbf{R_a}}  
   s_b \left( \rho_a \cdot \sigma -\frac{i Q(1-\Delta_a )}{2} \right)} ,\NN\\ 
&& s_b (z)
= \prod_{m=0}^\infty \prod_{n=0}^\infty
\frac{mb +nb^{-1}+Q/2 -iz}{mb +nb^{-1}+Q/2 +iz} .
\end{\eqa}
The basic idea for finding the Borel transformation is quite simple \cite{Honda:2016mvg}.
First we take polar coordinate for each gauge group $G_p$: $\sigma_i^{(p)} = \sqrt{\tau_p} \hat{x}_i^{(p)}$
with $\tau_p =i{\rm sgn}(k_p) t_p$ and $\hat{x}_i^{(p)}\in S^{{\rm rank}(G_p )-1}$
\footnote{
If $G_p$ is $SU(N)$, we consider $U(N)$ with insertion of $\delta (\sum_{j=1}^N \sigma^{(p)}_j )$.
}.
Then the partition function takes 
the form
\begin{\eqa}
&&  Z_{S_b^3}
= \Biggl[  \prod_{p=1}^n  i\sgn (k_p)  \int_0^{-i\sgn (k_p) \infty}  dt_p  
 e^{-\frac{t_p }{g_p}}   \Biggr]   f(i\sgn (k)t ) , \NN\\
&&   f (\tau ) 
=  \frac{\prod_{p=1}^n   \tau_p^{\frac{{\rm dim}(G_p )}{2}-1}}{2^n}  
\int_{\rm spheres}  d\hat{x}\    h (\tau ,\hat{x}) ,\NN\\
&&  h (\tau ,\hat{x}) 
=   Z_{\rm 1loop}(\sigma )    \prod_{p=1}^n  \tau_p^{-\frac{{\rm dim}(G_p )-{\rm rank}(G_p )}{2}}
\Bigr|_{\sigma_i^{(p)} 
= \sqrt{\tau_p} \hat{x}_i^{(p)}} .   \NN\\
\end{\eqa}
This is similar to the form of Borel resummation \eqref{eq:Borel} for multiple couplings
and it is natural to ask 
whether the function $f(\tau )$ is related to Borel transformation.
We can actually prove that
$f(\tau )$ is 
related to the Borel transformation of the perturbative series by \cite{Honda:2016vmv} 
\begin{\eq}
 \Biggl[ \prod_{p=1}^n i\sgn (k_p ) \Biggr] f( \{ \tau_p \} ) 
= \mathcal{B}Z_{S_b^3} \left( \{ -i\sgn (k_p ) \tau_p \} \right) .
\end{\eq}

Now we compare analytic property of the Borel transformation 
with the CSS in the last section.
When gauge group is product of rank-1 gauge groups,
we can easily see this
since $f(\tau )$ is no longer integral representation.
For example,
let us consider
a $U(1)$ CS theory with CS level $k>0$
coupled to charge $\{ q_a \}$ chiral multiplets with $R$-charges $\{ \Delta_a \}$.
Borel transformation of perturbative expansion of $Z_{S_b^3}$ by $g=1/\pi k$ in this theory is
\begin{\eq}
\mathcal{B}Z_{S_b^3} (t)
= \frac{1}{2\sqrt{-it}\prod_{a=1}^{N_f} 
   s_b \left( q_a \sqrt{it} -\frac{i Q(1-\Delta_a )}{2} \right)} .
\end{\eq}
This has simple poles and zeroes at
\begin{\eqa}
&& t_{\rm pole}^{m,n} 
= -\frac{i}{q_a^2} \left( mb+nb^{-1} +\frac{Q}{2}\Delta_a \right)^2 ,\NN\\
&& t_{\rm zero}^{m,n} 
= -\frac{i}{q_a^2} \left( mb+nb^{-1} -\frac{Q(\Delta_a -2)}{2}  \right)^2 , 
\end{\eqa}
which are parameterized by two integers $m,n\in\mathbb{Z}_{\geq 0}$.
Let us compare these with actions of the CSS.
Although the theory has various terms in general,
only relevant term for us is
the SUSY CS term 
\begin{\eqa}
S_{\rm CS} 
&=& \frac{ik}{4\pi}\int d^3 x \sqrt{g} {\rm Tr} 
                  \Biggl[ \eps^{\mu\nu\rho} \left( A_\mu \del_\nu A_\rho +\frac{2i}{3}A_\mu A_\nu A_\rho \right) \NN\\
&&                            -\blam\lambda +2D\sigma  \Biggr] ,
\end{\eqa}
because the CSS have vanishing Yang-Mills and matter actions 
\footnote{
For their explicit forms, see (2.9) and (2.18) in \cite{Fujitsuka:2013fga}. 
}.
In this theory,
the bosonic and fermonic CSS associated with $a$-th chiral multiplet have
\begin{\eqa}
&& \sigma_{m,n}^{\rm bos}
=-\frac{i}{q_a} \left( mb +nb^{-1} +\frac{Q}{2}\Delta_a \right) ,\NN\\
&& \sigma_{m,n}^{\rm fer}
=+\frac{i}{q_a} \left( mb +nb^{-1} -\frac{Q(\Delta_a -2)}{2} \right) ,
\end{\eqa}
respectively.
Therefore their actions are given by
\begin{\eqa}
 S_{\rm bos}
&=& \frac{i\pi k}{q_a^2} \left( mb+nb^{-1} +\frac{Q}{2}\Delta_a \right)^2
= \frac{t_{\rm pole}^{m,n}}{g} ,\NN\\
 S_{\rm fer}
&=& \frac{i\pi k}{q_a^2} \left( mb+nb^{-1} -\frac{Q(\Delta_a -2)}{2}  \right)^2
= \frac{t_{\rm zero}^{m,n}}{g} .
\end{\eqa}
Thus
single bosonic (fermionic) CSS 
with the action $S=S_c /g$
gives a simple pole (zero) of the Borel transformation at $t=S_c$.

For general rank, it is more involved
since there are integrals for $f(\tau )$
and not easy to find precise locations
of poles and zeroes of the Borel transformation.
However,
it is always true that
origins of all Borel singularities are the CSS
because all the singularities are technically coming from the poles of the one-loop determinant
whose locations are the same as the values of $\sigma$ on the CSS.
Although we have focused on the partition function,
the same result holds also for SUSY Wilson loops \cite{Tanaka:2012nr}
because its effect on Borel transformation is insertion of a function without poles and zeroes into $h(\tau ,\hat{x})$ \cite{Honda:2016vmv}.

\subsection{Round sphere limit}
For $b=1$, 
the building block $s_b (z)$ of the one-loop determinant
becomes a product over single integer \cite{Kapustin:2009kz}:
\begin{\eq}
 s_1 (z)
= \prod_{n=1}^\infty \left( \frac{n-iz}{n +iz } \right)^n .
\end{\eq}
Because of this,
poles and zeroes of Borel transformation become degenerate.
For example
in the $U(1)$ CS theory considered for the squashed case,
the Borel transformations have the poles and zeroes with degree-$n$ at
\begin{\eq}
t_{\rm pole}^{n} 
= -\frac{i( n +\Delta_a -1 )^2}{q_a^2}  ,\quad
 t_{\rm zero}^{n} 
= -\frac{i( n-\Delta_a +1 )^2}{q_a^2}  , 
\end{\eq}
with $n\in\mathbb{Z}_+$.
Hence the poles and zeroes parametrized by the two integers for general $b$
become coincident and are described by single integer for $b=1$.
This degeneration is reflected as the change of the degrees of the poles and zeroes.

Correspondingly,
the eigenvalues \eqref{eq:eigenB} and \eqref{eq:eigenF}
become degenerate for $b\rightarrow 1$
and
the CSS with the same $m+n$ have the same actions.
Therefore
we have 
$n$ bosonic and fermionic solutions with the actions
\begin{\eq}
 S_{\rm bos}
= \frac{t_{\rm pole}^{n}}{g} ,\quad
 S_{\rm fer}
= \frac{t_{\rm zero}^{n}}{g} ,
\end{\eq}
respectively.
Thus
the $n$-bosonic (fermionic) solutions 
with the coincident action $S=S_c /g$
give a degree-$n$ pole (zero) of the Borel transformation at $t=S_c$.
The same result holds also 
for various observables studied in \cite{Honda:2016vmv}.
Especially,
for superconformal case,
one can obtain the same conclusion 
for Bremsstrahrung function \cite{Lewkowycz:2013laa}, 
two point functions \cite{Closset:2012vg}
of stress tensor and $U(1)$ flavor symmetry current on $\mathbb{R}^3$.
Therefore 
we expect that
for superconformal case (not limited to 3d $\mathcal{N}=2$ theory)
CSS on sphere provide information on analytic property of Borel transformation even for flat space case.

\subsection{Hyper multiplets on round sphere}
When matters are only hyper multiplets,
the one-loop determinant becomes much simpler:
\begin{\eq}
 Z_{\rm 1loop}(\sigma ) 
= \frac{\prod_{\alpha \in {\rm root}_+ }
 4\sinh^2{(\pi \alpha \cdot \sigma )}  }
   {\prod_a \prod_{\rho_a \in \mathbf{R_a}}  
   2\cosh{(\pi \rho_a \cdot \sigma )}} ,
\end{\eq}
because of the identity 
\begin{\eq}
\frac{1}{s_1 \left( z -i/2 \right) s_1 \left( -z -i/2 \right)} 
= \frac{1}{2\cosh{(\pi z)} }  . 
\end{\eq}
This simplification comes 
from cancellations of bosonic and fermionic contributions 
in a pair of conjugate representations with the $R$-charges $1/2$.

For example,
if we consider a $U(1)$ CS theory with CS level $k>0$
coupled to charge $\{ q_a \}$ hyper multiplets,
then the Borel transformation does not have zeroes
but has simple poles at
\begin{\eq}
t_{\rm pole}^{n} 
= -\frac{i(2n-1)^2}{4q_a^2}  ,\quad n\in\mathbb{Z}_+ .
\end{\eq}
Correspondingly 
we have $n_B$ bosonic and $n_F$ fermionic solutions
with the following actions
\begin{\eq}
 S_{\rm bos}
= \frac{i(2n_B -1)^2}{4q_a^2 g}  ,\quad
 S_{\rm fer}
= \frac{i(2n_F +1)^2}{4q_a^2 g} ,
\end{\eq}
with $n_B ,n_F \in\mathbb{Z}_+$.
For solutions with the action $t_{\rm pole}^{n} /g$,
we have $n$ bosonic and $(n-1)$ fermionic solutions
and hence always $n_B -n_F =1$ for any $n$.
Thus the CSS predict that
the degree-$n$ poles at $t=t_{\rm pole}^{n} $ are canceled by the degree $(n-1)$-zeroes and becomes the simple poles.
This agrees with our conjecture. 

\subsection{Planar limit}
It is usually expected that
the perturbative series by 't Hooft coupling in the planar limit
is convergent \cite{'tHooft:1982tz}
as the number of Feynman diagrams
do not grow as factorial.
Hence,
if there are no renormalons,
we expect that
Borel transformation in the planar limit 
does not have singularities.
We can easily see that
our CSS are indeed consistent with this expectation.
In the planar limit $N\rightarrow\infty$ with fixed $N/k$,
the actions of the CSS become infinite
unless we have additional factor of order $1/N$.
Thus in typical ``matrix type" theory
like
large-$N$ CS theory 
coupled to chiral multiplets with up to two index representations,
Borel singularities go away 
and we do not have Borel singularities in the planar limit.

\section{Conclusion and discussions}
In this paper
we have pointed out physical importance of complexified supersymmetric solutions
which satisfy SUSY conditions but are not on original path integral contour.
We have conjectured that
if there are $n_B$ bosonic and $n_F$ fermionic CSS
with the action $S=S_c /g$,
then Borel transformation 
has a pole for $n_B \geq n_F$ and zero for $n_B \leq n_F$ with degree $|n_B -n_F |$ at $t=S_c$.
We have shown this statement
for various SUSY observables 
in the 3d $\mathcal{N}=2$ SUSY CS matter theories on sphere.
We have constructed the CSS 
in general Langrangian 3d $\mathcal{N}=2$ theory
and compared their actions with analytic property of the Borel transformations
to see agreement with our conjecture.

We have seen that
the fermionic CSS give information on the zeroes of the Borel transformation.
To our knowledge
physical implication of Borel zeroes have not been explored in literature.
Naively Borel zeroes do not seem to be related 
to large order behavior of perturbative series unless they are ends of branch cuts.
There might be general argument 
for a relation between Borel zeroes and fermionic saddle points
as for the relation between Borel singularities and saddle points \cite{Lipatov:1976ny,Argyres:2012vv}.

We have not discussed
whether the CSS are contributing saddle points or not.
In order to answer this question,
it is appropriate to perform Lefschetz thimble analysis for 3d $\mathcal{N}=2$ theories. 
More generally,
since most of the analyzes of SUSY localization in literature have not performed serious complex saddle point analysis 
to our knowledge,
there is a logical possibility that
we are missing contributions from complex saddles in most setups.
Perhaps the setups in \cite{Fujitsuka:2013fga,Benini:2013yva,Benini:2015noa}
would be good starting points to address this issue.

It is known that
in the (Coulomb branch) localization formula for $Z_{S_b^3}$,
picking up poles of the one-loop determinant in a half complex plane
gives rise to Higgs branch representation of the partition function
including a product of vortex and anti-vortex partition functions 
for some theories \cite{Pasquetti:2011fj,Fujitsuka:2013fga,Benini:2013yva,Alday:2013lba}.  
Since we have seen the correspondence between the poles and bosonic CSS,
presumably the CSS are related to the Higgs branch formula or vortices.
It is interesting to make this intuition more precise.

Recently there appeared multi-cut solutions 
in matrix models obtained by localization of the $S^3$ partition function of 3d $\mathcal{N}=2$ CS theories \cite{Morita:2017oev}.
In the solutions,
poles and zeroes in the one-loop determinant play important roles
to gather eigenvalues of the matrix models.
It would be illuminating to study connection between the solutions and our CSS.

\subsection*{Acknowledgements}
A part of this work was presented 
in ``Geometry of String and Gauge Theories" at CERN,
``RIMS-iTHEMS International Workshop on Resurgence Theory" at RIKEN
and lectures at Fudan University.
The author is grateful to 
B.~Assel, C.~Closset, S.~Cremonesi, T.~Fujimori, K.~Hosomichi, S.~Kamata, 
T.~Misumi, T.~Morita, M.~Nitta and N.~Sakai 
for useful discussions.
The author would like to thank 
Centro de Ciencias de Benasque,
CERN, Fudan University, KITP, 
National Taiwan University, RIKEN and YITP for hospitalities.


\end{document}